\documentclass[article,12pt]{elsarticle}
\usepackage{amsmath,amsfonts,amssymb,amsthm,epsfig,subfigure,graphicx,url}
\usepackage{cancel}
\usepackage{eufrak}
\usepackage[utf8]{inputenc}
\usepackage{color}
\usepackage{graphicx}
\usepackage{subcaption}

\renewcommand{\ge}{\geqslant}

\newcommand{\be}{\begin{equation}}
\newcommand{\en}{\end{equation}}

\newcommand{\bea}{\begin{eqnarray}}
\newcommand{\eea}{\end{eqnarray}}
\newcommand{\beas}{\begin{eqnarray*}}
\newcommand{\eeas}{\end{eqnarray*}}

\usepackage{color}

\newcommand{\cn}{\color{black}}

\usepackage{geometry}
\begin{document}
\numberwithin{equation}{section}

\begin{frontmatter}

\title{An analytic study on the properties of solitary waves traveling on tensegrity-like lattices}

\author{Ada Amendola}

\address{Department of Civil Engineering, University of Salerno, Via Giovanni Paolo II, 132 - 84084 Fisciano (SA), Italy. 
E-mail:
adaamendola1@unisa.it}


\begin{abstract}
This paper develops an analytic study on the existence and properties of solitary waves on 1D chains of lumped masses and nonlinear springs, which exhibit a mechanical response similar to that of tensegrity prisms with locking-type response under axial loading. Making use of the Weierstrass’ theory of 1D Lagrangian conservative systems, we show that such waves exist and that their shapes depend on the wave speed. A progressive localization of the traveling pulses in narrow regions of space is observed as the wave speed increases up to a limit value. A comparative analysis illustrates that the presented study is able to capture the wave dynamics  observed in previous numerical studies on tensegrity mass-spring chains.

\end{abstract}


\end{frontmatter}



\section{Introduction}

The use of the tensegrity paradigm to design novel mechanical metamaterials offers several key advantages beyond conventional systems \cite{Skelton}. 
This is mainly due to the possibility of designing systems with all members axially loaded (allowing for a global bending without member bending); structural efficiency in terms of remarkably high stiffness and strength-to-mass ratios; a marked tunability because the stiffness of the structure can be easily controlled through internal/external prestress; high scalability (size-independent properties); and the ubiquity in nature of tensegrity systems, which permits one to design nature-inspired systems \cite{Skelton}.
Although several slightly different definitions of the terms `acoustic’ and `mechanical metamaterials’ are found in the literature, they all broadly refer to engineered materials exhibiting unconventional properties deriving from the arrangement of their internal structure rather than from single constituents \cite{Lu,Kadic}. Typically, a metamaterial is a periodic structure composed of repeated building blocks (or “meta-atoms”) that form systems exhibiting exceptional static and dynamic behaviors.
Nonlinear metamaterials react to strains, stresses, and mechanical energy inputs in “exotic” fashion through shape-morphing functionalities and large displacements \cite{Bertoldi}. 
Recent research has shown that tensegrity lattices (formed, e.g., by tensegrity prism units) act like nonlinear springs that can switch their response from softening to stiffening in the large displacement regime \cite{Opp1}-\cite{[30]}.
The stiffening response is activated, e.g., when adjacent cells of a tensegrity lattice are pushed against each other, or the units are close to a “locking” configuration characterized by an infinitely large (tangent) axial stiffness. Such a behavior supports the formation of solitary waves localized in narrow regions of space within tensegrity mass-spring chains \cite{[24],[27],[30]}, which is useful for designing acoustic lenses with  advanced wave‐focusing features (see, e.g., \cite{[75],[76],[77]} and references therein).

This study deals with an analytic study on the characterization of the (quasicontinuum) limit and the wave dynamics of a 1D metamaterial formed by nonlinear springs with tensegrity-like response, which alternate with lumped masses. 
The analyzed spring-units exhibit an elastic potential that retains the basic properties of that ruling the response of tensegrity prisms equipped with rigid bases, i.e., a locking response in compression \cite{[24]}. We obtain analytic results on the shape and properties of the propagating pulses, using the Weierstrass’ theory of 1D Lagrangian conservative systems \cite{DGS}, \cite{Cirillo}. We analytically prove the existence of solitary pulses that exhibit a localization behavior for increasing values of the wave speed, 
up to a limit value of such a quantity. The results provide, for the first time, an analytic characterization of the pulses that travel across 1D tensegrity-like lattices with hardening response, since the existence and properties of such waves have been studied only through numerical approaches thus far \cite{[24],[27]}.
  Analytic results have been obtained for Hertzian chains, but with reference to different, compacton-type solitary waves \cite{Nesterenko,Anco17,Anco18}. 
\cn
The paper is organized as follows.
Section \ref{sectV} presents a nonlinear, \textit{tensegrity-like} potential with locking-response under compression loading. Section \ref{secqc} deals with the quasicontinuum limit of the equations of motion of a compressed chain of lumped masses alternating with tensegrity-like springs.
The existence and properties of solitary waves traveling across such a system are discussed in Sect. \ref{secsolitary}. This section also presents numerical comparisons between the presented theory and available results on the wave dynamics of tensegrity mass-spring chains \cite{[24]}.
The work ends in Sect. \ref{conclusions} with concluding remarks and suggestions for future research.


\section{Tensegrity-like mass spring chains}
\label{sectV}

\begin{figure}[h!] 
\centering	
\includegraphics[width=0.65\textwidth]{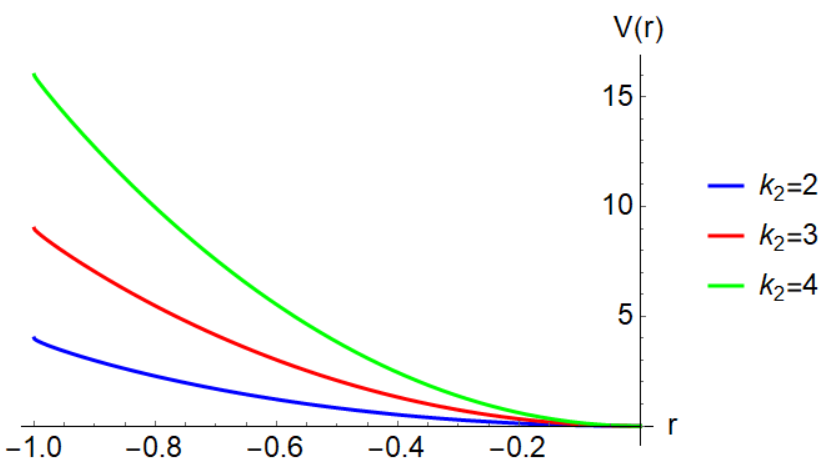} \\
\includegraphics[width=0.65\textwidth]{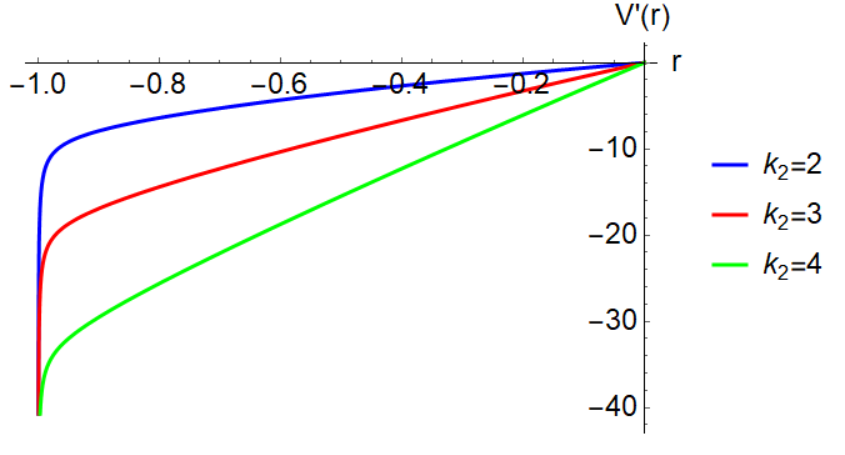}
\caption{Plots of the $V(r)$ and $V'(r)$ functions for $k_1=1$, $r_{lock}=-1$, and different values of $k_2$.} \label{fig1}
\end{figure}

Let us consider a one-dimensional mass-spring chain formed by lumped masses and nonlinear springs loaded in compression. The generic spring features rest length $h_0$ and deformed length $h$, with $h_0 \ge h \ge h_{lock}$.
Here, $h_{lock}$ denotes the length at which the springs get \textit{locked}, i.e., its length cannot be further reduced.
The spring under examination is equipped with the following elastic potential:
\be \label{Vpotr}
V(r)=-k_1 r^2 \left(\sqrt{r-r_{lock}}-k_2^2\right),
\en 
where $r=h-h_0$, $r_{lock}=h_{lock}-h_0$, and $k_1$ and $k_2$ are constitutive parameters. The first derivative of potential  \eqref{Vpotr} is given by
\be \label{DVpotr}
V'(r)=-2 k_1  r
   \left(\sqrt{r-r_{lock}}-k_2^2 \right) -\frac{
   k_1 r^2}{2 \sqrt{r-r_{lock}}},
\en 
and such a function shows a vertical asymptote for $r = r_{lock}$.
Fig. \ref{fig1} plots the functions $V(r)$ and $V'(r)$ assuming $r_{lock}=-1$,  $k_1=1$, and different values of $k_2$.

Let us now set $u=-r$,  $u_{lock}=-r_{lock}=h_0-h_{lock}$, and $\xi=u/u_{lock}$. Under such settings, the potential \eqref{Vpotr} can be rewritten as:
\be \label{eqVpotxi} 
V(\xi)=-\hat{k}_1 \xi^2 \left(\sqrt{1-\xi}-\hat{k}_2^2\right),
\en
\noindent with 
\begin{eqnarray} \label{ktilde} 
\hat{k}_1 & = \ k_1 \ u_{lock}^{\frac{5}{2}}, \\ 
\hat{k}_2^2 & = \ k_2^2 \  u_{lock}^{-\frac{1}{2}}
\end{eqnarray}



\begin{figure*}[tbh]
    \centering	
    \includegraphics[scale=0.47]{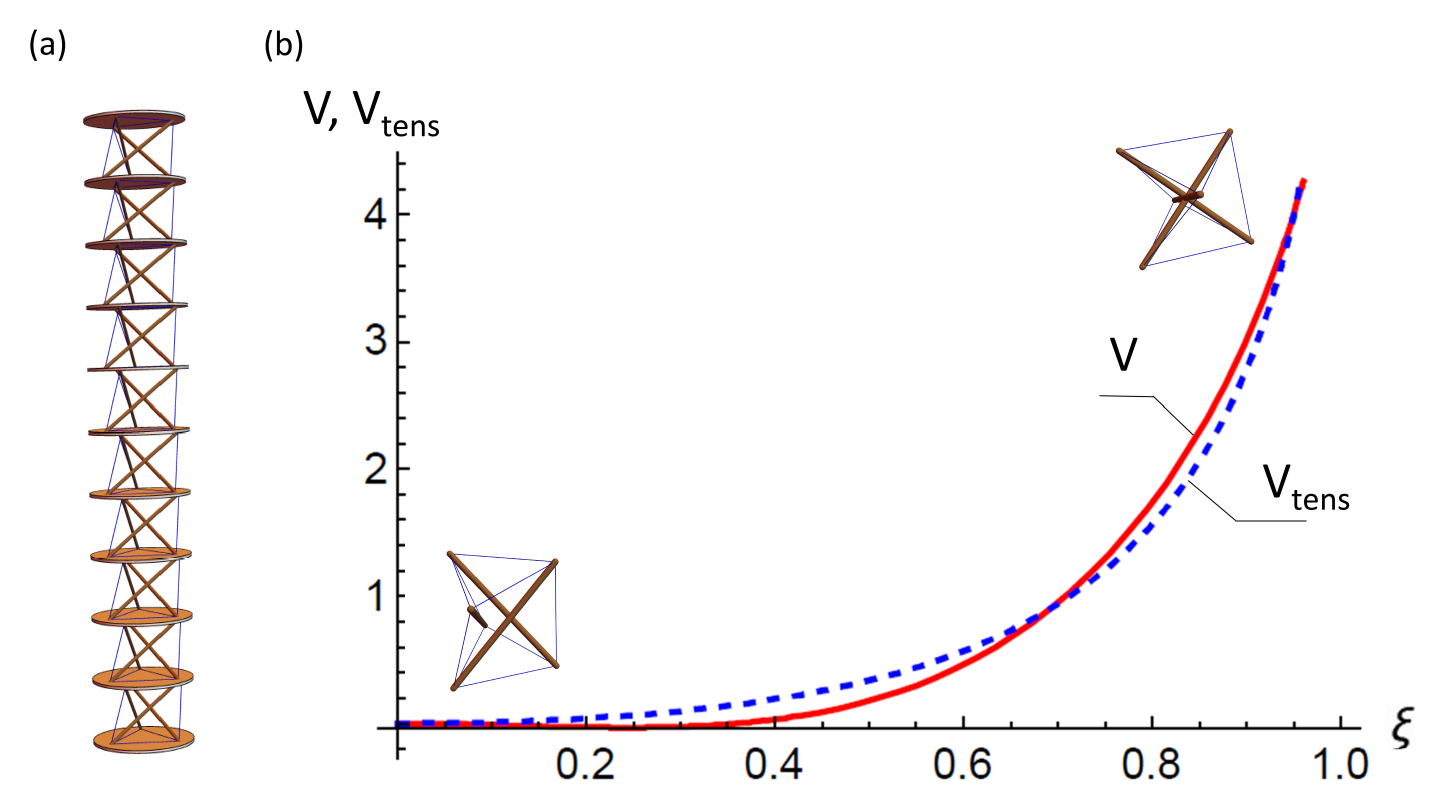} \\
    \caption{{(a)} Tensegrity mass-spring chain. {(b)} Fitting of the tensegrity potential $V_{tens}(\xi)$ analyzed in Ref. \cite{[24]} (blue-dashed curve) with the tensegrity-like potential \eqref{eqVpotxi}  (red-solid curve).
    }
    \label{T-Tlike}
\end{figure*}

It is worth noting that this potential captures the locking property in compression of that characterizing the tensegrity mass-spring chains with hardening response analyzed in \cite{[24],[27]}.  Such chains are formed by 1D arrays of lumped masses and minimal regular tensegrity prisms \cite{Skelton} (Fig. \ref{T-Tlike}{(a)}). The $i$-th prism is built using  two discs (the end faces), three uniform rigid bars with fixed length $L$, and three diagonal cross-cables of identical length $\lambda$ and linearly elastic stiffness $k$, connecting the nodes of the triangular bases.
The assumption of frictionless contact between the prisms and the discs ensures that that no bending moments and torques are transmitted from the prisms to the discs during the motion of the system. The additional assumption that the tensegrity prims are much lighter than the spacing discs allows one to lump the overall mass $m$ of the generic unit (including the mass of a prism and that of a disc) in correspondence of the current disc.
The potential $U(r)$ that describes the tensegrity chain analyzed in Ref. \cite{[24]} can be written as a function of $\xi$ in this form:

\bea
U_{tens}(\xi) & = & \frac{3}{2} \ k  \left[\frac{ \sqrt{3 (h_0-\xi u_{lock})^2 - L^2 + a^2(6 - \sqrt{3}c(\xi))} }{\sqrt{2}} \ - \ \frac{\lambda_0}{1+p} \right]^2 
\label{Ur}
\eea

\noindent where $L$ is the length of the bars; $a$ is the radius of the circumcircles of the triangular bases (the bases of the prisms, as well as the bars, are assumed to transform rigidly during the motion of the prism);  $\lambda_0$ is the length of the cross-strings in correspondence to the self-standing configuration (possibly internally prestressed); and $p$ is the prestrain of the cross-strings. The function $c(\xi)$ appearing in Eqn. \eqref{Ur} is given by 

\bea
c(\xi) & = & \sqrt{ \frac{[L^2 - (h_0-\xi u_{lock})^2 ] \ [4 a^2 + (h_0 -\xi u_{lock})^2 - L^2]}{a^4} }
\label{cr}
\eea

Upon setting  $V_{tens}(\xi)=U_{tens}(\xi)-U_{tens}(0)$, and assuming the properties of the physical model of a tensegrity mass-spring chain analyzed in \cite{[24]} ($L = 0.180$ m; $a=0.070$ m; $h_0=0.119$m; $k=4.6417 \times 10^4$ N/m; $p=2\%$; $u_{lock}=5.66$ mm), we employed the Fit function of Mathematica$^{\makebox{\textregistered}}$ (v. 12.2)
to approximate  $V_{tens}$ through the function \eqref{Vpotr}. We obtained the best fit parameters: 
${\hat k}_1=7.72021$, ${\hat k}_2=0.893722$ 
(Fig. \ref{T-Tlike}(b)). It is worth observing that the locking configuration of the prisms forming the tensegrity chain is achieved when the three bars touch each other (see the inset to the top right of (b) panel in Fig. \ref{T-Tlike}).

The remainder of the paper uses the nomenclature “tensegrity-like” for the potential \eqref{Vpotr}. Such a potential is easier to handle than the $V_{tens}$ potential from the analytic point of view, and will be employed to develop the wave dynamics study presented in the next sections. The results given in Sect. \ref{secsolitary} will show that the shapes of the pulses traveling in tensegrity-like chains are similar to those of the pulses that traverse the tensegrity chain studied in Ref. \cite{[24]}.

\section{Quasi-continuum limit of the equations of motion} \label{secqc}
We now consider the \emph{quasi-continuum} limit of a tensegrity-like mass spring chain. 
In the absence of external forces, the equation of motion of the $i$-th mass is written as follows: 

\be \label{eq1}
m\ddot{u}_{i}=V_{u}(u_{i-1}-u_i)-V_{u}(u_i-u_{i+1}),
\en
where $V_u$ is the derivative of the tensegrity-like potential $V$ with respect to $u$, while $u_{i-1}$, $u_{i}$, and $u_{i+}$ are the displacements exhibited by the masses $i-1$, $i$, and $i+1$, respectively. 
Let $x$ denote a longitudinal coordinate measured along the chain, and let $x_i$ denote the value of $x$ at the $i$-th mass. On considering a chain with infinite particles, expressing the displacement of the $i$-th mass as the sample value of a continuous function $u(t; x)$ at $x=x_i$, and making use of the quasi-continuum approach presented in \cite{quasicont,Olver}, we are led to transform Eqn. \eqref{eq1} into this \textit{Boussinesq equation}:

\be \label{eq3}
m u_{tt}=[V_u]_{xx}+\gamma u_{xxxx},
\en

\noindent where $\gamma$ is a quantity that accounts for terms of order higher than two in the Taylor expansion of $u$ \cite{Olver}.
From \eqref{eq3}, via a simple manipulation, we get

\be \label{eq4bis}
\xi_{tt}= \left[ \tilde{V}_\xi \right]_{xx}+ \gamma \xi_{xxxx}.
\en
where $\xi(t; x)=u(t; x)/u_{lock}$; $\tilde{V}(\xi)={V}(\xi)/(m  u_{lock}^2)$;
${\tilde{k}_1}={\hat{k}_1}/(m  u_{lock}^2)={k_1 u_{lock}^{1/2}}/{m}$;
${\tilde{k}_2}={\hat{k}_2}$; and

\be \label{eqVpotxim} 
\tilde{V}(\xi)=-{\tilde{k}_1} \xi^2 \left(\sqrt{1-\xi}-\tilde{k}_2^2\right).
\en

\indent We now consider a traveling $\xi$-wave with speed $v$. By setting  $\xi=\Phi(x-v t)$ into \eqref{eq4bis}, and adopting the prime notation for the differentiation with respect to $z=x-v t$, we obtain
\be \label{eq4ter}
v^2 \Phi^{''} \ = \ 
[ \tilde{V}_{\Phi} ]^{''} \ + \ \gamma  \Phi^{''''}.
\en
A double integration of such an equation with respect to $z$ and the adoption of asymptotic boundary conditions to eliminate the integration’s constants leads us to
\be \label{eqtmp}
{v}^2 \Phi \ = \ \tilde{V}_{\Phi} + \gamma \Phi^{''}.
\en
We now multiply both sides of \eqref{eqtmp} by $\Phi^{'}$, and we take an additional integration with respect to $z$, to finally obtain
\be \label{equu}
\mathcal{F}  \ = \ {v}^2 \Phi^2 \ - \ 2 \tilde{V} + C,
\en
where $\mathcal{F} = \gamma \Phi^{' 2}$ and $C$ is the integration constant. 

\section{Existence and properties of solitary waves} \label{secsolitary}

Unpon inserting the expression \eqref{eqVpotxi} of the tensegrity-like potential 
into Eqn. \eqref{equu}, setting the integration constant $C$ equal to zero \cite{Peli1}, we get to the equation

\be \label{eqFcal}
\mathcal{F} \ = \ \Phi^2 \left[{v}^2 +2{\tilde{k}_1} ( \sqrt{1-\Phi}-\tilde{k}_2^2) \right].
\en

The Weierstrass discussion of one degree of freedom equations can be usefully employed to detect whether Eqn. \eqref{eqFcal} admits pulse solitary waves as solutions, which correspond to wave profiles featuring an asymptotic point followed by an inversion point (see, e.g., \cite{DGS} and references therein). $\mathcal{F}$ exhibits a zero of degree two for $\Phi=0$, which matches an asymptotic point of the $\xi$-pulse \cite{DGS}. An additional simple zero of $\mathcal{F}$ is found at the point $\Phi^* \in ]0,1]$ where the term in square brackets of \eqref{eqFcal} is zero. This results in

\be \label{eqphistar}
\Phi^*=\frac{-v^4+4{\tilde{k}_1}^2 v^2 \tilde{k}_2 + 4 {\tilde{k}_1}^2\tilde{k}_2^2}{4\tilde{k}_2^2}.
\en
and such a point locates the peak of the pulse. By setting $\Phi^*=1$, one obtains this limit value of pulse speed: $v_{lim}=\tilde{k}_2 \sqrt{2\tilde{k}_1}$. It is worth noting that $\Phi^* \in ]0,1]$ is guaranteed for $v\in[0,v_{lim}]$. 
The shape of the solitary pulse is obtained in implicit form as 

\be \label{eqFcal1}
f \ := \ \int \frac{d\Phi}{\sqrt{\mathcal{F}}}=\pm \left(\gamma^{-1}z+\, constant\,\right),
\en
\noindent where the arbitrary constant is of no interest because it is only a way to center the pulse. 
The cumbersome expression of function $f$ in \eqref{eqFcal1} is as follows
\be \label{ffunc}
f \ = \ \frac{1}{\chi \Phi} \ - \ \frac{2\chi \ \Phi \ \rho \ \sqrt{\psi} \ \arctan[\frac{\chi}{\psi}] \ + \ \sqrt{\rho} \ \psi \ \arctan[\frac{\chi}{\rho}]}{ \rho \ \ \psi \chi}
\en
\noindent where
\bea \label{ffuncparam}
\chi & = \  \sqrt{v^{2}+2k_1(-k_2^2+\sqrt{1-\Phi})}, \nonumber \\
\rho & = \ (v^2-2 (k_2^2 - 1) k_{1}),  \ \ \ \  \psi \ = \ (v^2-2 (k_2^2 + 1) k_{1}) . 
\eea

\begin{figure}[ht] 
\centering	
\includegraphics[width=0.70\textwidth]{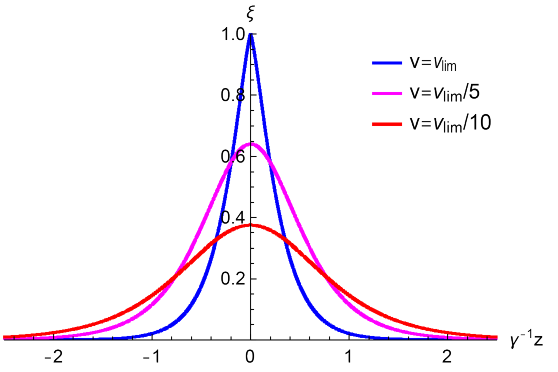}
\caption{Strain wave profiles in tensegrity chains for different wave speeds \cite{[24]}, compared with those obtained through the tensegrity-like potential (we assumed ${\hat k}_1=7.72021$, ${\hat k}_2=0.893722$, $\gamma=100$).} \label{fig2}
\end{figure}

\begin{figure}[htp] 
\centering	
\includegraphics[width=0.85\textwidth]{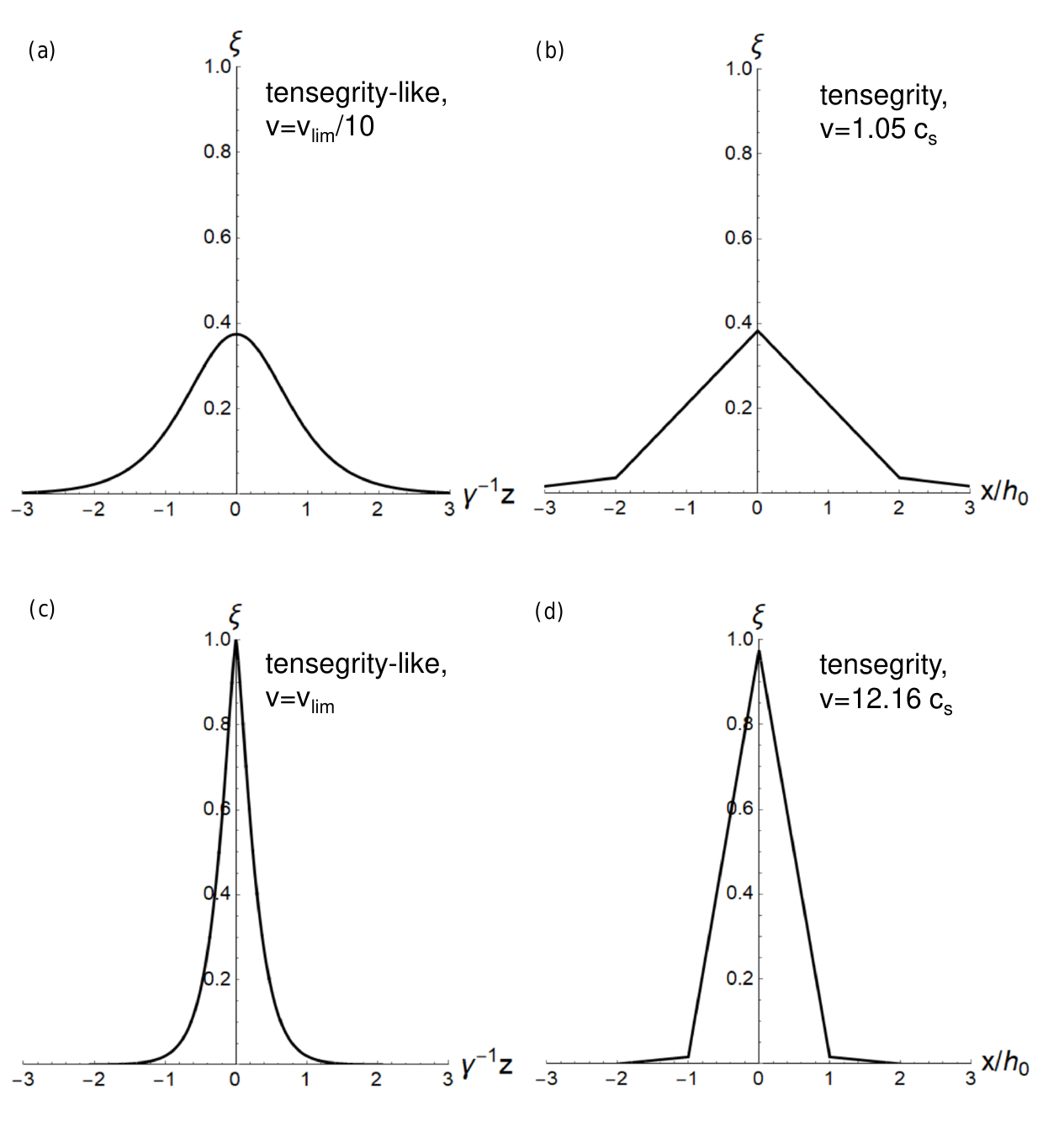}
\caption{Wave profiles in tensegrity chains for different wave speeds obtainded in \cite{[24]}, in comparison with analytical results obtained by employing the tensegrity-like potential (${\hat k}_1=7.72021$, ${\hat k}_2=0.893722$, $\gamma=100$).} \label{panel}
\end{figure}

Fig. \ref{fig2} shows the pulse profiles obtained using Eqn.\eqref{ffunc}, the constitutive parameters given in Sect. \ref{sectV}, and different values of $v$; $\gamma$ is fixed equal to $100$ to allow a consistent and direct comparison of the predictions of the tensegrity-like model with the numerical results presented in \cite{[24]}. 
The results in Fig. \ref{fig2} show that the solitary pulses traveling   in tensegrity-like chains localize in narrow regions of space as $v$ approaches $v_{lim}$, tending to assume a \textit{peakon-like} profile for $v \rightarrow v_{lim}$. \cn
The wave propagation regime with $v>v_{lim}$ is not accessible to the system, since it locally reaches a locking condition for $v=v_{lim}$ ($\xi=1$, see Figs. \ref{fig2}-\ref{panel}), and additional dynamic inputs 
and additional dynamic inputs would would not be able to propagate.

Fig. \ref{panel} shows a comparison between the shapes of the pulses traveling in tensegrity-like and tensegrity mass-spring chains. The results for a tensegrity mass-spring chain are imported from Ref. \cite{[24]}, considering two different values of the ratio between the speed of the pulse $v$ and the linearized speed of sound of the system $c_s$ (we examine a nearly sonic pulse with $v 1.05 c_s$, and a supersonic pulse with $v=12.16 c_s$ ). One observes a good qualitative match between the shapes of the examined wave profiles, which both exhibit a localization behavior for increasing values of the wave speed.


\section{Concluding Remarks} \label{conclusions}

This work provides an analytic characterization of the solitary pulses that propagate in tensegrity-like lattices, which are obtained by alternating lumped masses and nonlinear springs with locking behavior in compression. The shape of the pulses traveling through such systems has been derived in implicit form using the Weierstrass theory of one-dimensional Lagrangian systems.
Previous results on the localization of solitary pulses in the supersonic wave regime of tensegrity mass-spring chains have been confirmed by this study. Such peculiar dynamic behavior can be usefully exploited to design mechanical metamaterials with enhanced wave‐focusing capabilities, as well as a radically new actuators/sensors technology for the generation of stress waves and damage detection \cite{[75],[76],[77]}. We recommend this exciting topic be explored in future work. Additional lines of future research will include an investigation of the stability of the solitary wave solutions obtained in this work, as well as an analytic study on 
rarefaction solitary waves traveling in tensegrity-like mass spring chains with softening-type response \cite{[27]}.

\vskip0.5cm
\noindent \textbf{ACKNOWLEDGEMENTS.}  
The author is grateful to the Italian Ministry of University and Research for support through the PRIN 2017 grant 2017J4EAYB. The present work has been carried out under the auspices of the Italian `Gruppo Nazionale per la Fisica Matematica' (GNFM) of the `Istituto Nazionale di Alta Matematica Francesco Severi' (INDAM).


\end{document}